\documentclass[twocolumn,showpacs,amsmath,amssymb,10pt]{revtex4}
\usepackage{graphicx,color}
\usepackage{amsmath}
\usepackage{amssymb}
\usepackage{bbm}

\usepackage{amsfonts}
\usepackage{bm}

\newcommand{\ot}{{\,\otimes\,}}
\newcommand{{\Cd}}{{\mathbb{C}^d}}

\newcommand{\tr}{\mathrm{Tr}}
\newcommand{\bra}[1]{\mbox{$\langle #1 |$}}

\newcommand{\ket}[1]{\mbox{$| #1 \rangle$}}
\def\oper{{\mathchoice{\rm 1\mskip-4mu l}{\rm 1\mskip-4mu l}
{\rm 1\mskip-4.5mu l}{\rm 1\mskip-5mu l}}}
\def\<{\langle}
\def\>{\rangle}

\newtheorem{Example}{Example}
\newtheorem{Remark}{Remark}

\newcommand{\beq}{\begin{equation}}
\newcommand{\eeq}{\end{equation}}
\newcommand{\bear}{\begin{eqnarray}}
\newcommand{\ear}{\end{eqnarray}}
\newcommand{\bdm}{\begin{displaymath}}
\newcommand{\edm}{\end{displaymath}}

\begin{document}

\title{\textbf{%On Markovianity of random unitary quantum evolution \\
Non-Markovianity degree for random unitary evolution}}
%
%From Markovian to maximally non-Markovian evolution in random unitary dynamics}}
\author{Dariusz Chru\'sci\'nski and Filip A. Wudarski }
\affiliation{ Institute of Physics, Faculty of Physics, Astronomy and Informatics \\  Nicolaus Copernicus University,
Grudzi\c{a}dzka 5/7, 87--100 Toru\'n, Poland
}

\begin{abstract}
We analyze the non-Markovianity degree for random unitary evolution of $d$-level quantum systems. It is shown how
non-Markovianity degree is characterized in terms of local decoherence rates. In particular we derive a sufficient condition for vanishing of the backflow of information.

\end{abstract}

\pacs{03.65.Yz, 03.65.Ta, 42.50.Lc}

\maketitle

%======SECTION=================INTRODUCTION=================

%\section{Introduction}

Recently, much effort was devoted to {the}  analysis  of non-Markovian quantum evolution \cite{R1}--\cite{Yu} (see also \cite{RIVAS} for the recent review).
%In particular various concepts of non-Markovianity were introduced and several so called non-Markovianity measures were proposed.
The two most popular  approaches  are based on divisibility of the corresponding dynamical map \cite{RHP,Wolf-Isert,Hou} and distinguishability of states  \cite{BLP}. Other approaches use quantum entanglement
\cite{RHP}, quantum Fisher information \cite{Fisher}, fidelity \cite{fidelity}, mutual information \cite{Luo1,Luo2}, channel capacity \cite{Bogna,Carole},  geometry of the set of accessible states \cite{Pater}, non-Markovianity degree \cite{PRL-Sabrina} and the quantum regression theorem \cite{Petruccione,Vacchini-14}. There is also an alternative approach based on the so called Di\'osi-Gisin-Strunz non-Markovian quantum state diffusion equation \cite{Yu} but we do not consider it in this paper.

In what follows we analyze non-Markovianity degree of random unitary quantum evolution of $d$-level quantum system. Let us briefly recall the notion of non-Markovianity degree \cite{PRL-Sabrina}: if $\Lambda_t$ is a dynamical map then it is called $k$-divisible iff the corresponding propagator $V_{t,s}$ defined {\em via} $\Lambda_t = V_{t,s} \Lambda_s$ ($t \geq  s)$ defines $k$-positive map \cite{k-pos}. Hence, if the system Hilbert space is $d$-dimensional, then $k\in \{1,2,\ldots,d\}$. Map which is $d$-divisible we call CP-divisible (the corresponding propagator is completely positive (CP)) and $1$-divisible we call P-divisible  (the corresponding propagator is  positive (P)). The evolution is Markovian iff the corresponding dynamical map is CP-divisible. Note that if $\Lambda_t$ is $k$-divisible, then is is necessarily $l$-divisible for all $l< k$. Maps which are even not P-divisible we call essentially non-Markovian. Having defined the notion of $k$-divisibility one assign the non-Markovianity degree as follows: if $\Lambda_t$ is $k$-divisible (but not (k+1)-divisible), then its non-Markovianity degree ${\rm NMD}[\Lambda_t] = d-k$. Clearly, if $\Lambda_t$ is Markovian, then ${\rm NMD}[\Lambda_t] = 0$ and if  $\Lambda_t$ is essentially non-Markovian, then ${\rm NMD}[\Lambda_t] = d$.

Let us recall that a  quantum channel $\mathcal{E} : \mathcal{B}(\mathcal{H}) \rightarrow \mathcal{B}(\mathcal{H})$ is called random unitary if its Kraus representation is given by
\begin{equation}\label{}
  \mathcal{E}(X) = \sum_k p_k\, U_k XU_k^\dagger\ ,
\end{equation}
where $U_k$ is a collection of unitary operators and $p_k$ stands for a probability distribution. The characteristic feature of such channels is unitality, that is, $\mathcal{E}(\mathbb{I}) = \mathbb{I}$. Actually, for qubits (${\rm dim}\, \mathcal{H}=2$), it turns out \cite{Landau} that any unital channel is random unitary. However, for higher level systems it is no longer true. A random unitary dynamics is represented by a dynamical map $\Lambda_t$ such that for all $t > 0$ the channel $\Lambda_t$ is random unitary.

%\section{Random unitary dynamics}

Consider the following set of unitary generalized spin (or Weyl) operators in $\mathbb{C}^d$ defined by
\beq
U_{kl}=\sum_{k,l=0}^{d-1} \omega^{kl}\ket{m}\bra{m+ l},
\eeq
with $\omega=e^{2\pi i/d}$. They satisfy well known relations
\beq
U_{kl}U_{rs}=\omega^{ks}U_{k + r,l + s},\quad U_{kl}^\dag=\omega^{kl}U_{-k,-l}.
\eeq
Introducing a single index $\alpha \equiv (m,n)$ via $\alpha = md +n$ ($\alpha=0,\ldots,d^2-1$). One has $U_0 = \mathbb{I}$ and $\tr[U_\alpha U_\beta^\dagger]= d \delta_{\alpha\beta}$ for $\alpha,\beta=0,1,\ldots,d^2-1$.
In this paper we consider a random unitary evolution defined by the following dynamical map
\beq
\Lambda_t(X)=\sum_{\alpha=0}^{d^2-1}\, p_{\alpha}(t)\, U_{\alpha}XU_{\alpha}^\dag\ ,
\eeq
with time-dependent probability distribution $p_\alpha(t)$ satisfying $p_0(0)=1$. Assuming time-local Master Equation
\begin{equation}\label{}
  \dot{\Lambda}_t = L_t \Lambda_t\ ,
\end{equation}
it is well known that $\Lambda_t$ is CP-divisible iff $L_t$ has the standard Lindblad form for all $t\geq 0$. To find the time-local generator $L_t$ let us observe that
\begin{equation}\label{}
  \Lambda_t(U_{\alpha})=\lambda_{\alpha}(t)U_{\alpha}\ ,
\end{equation}
where the eigenvalues $\lambda_\alpha(t)$ read as follows
\begin{equation}\label{}
  \lambda_{\alpha}(t) = \sum_{\alpha,\beta=0}^{d^2-1} H_{\alpha\beta} p_{\beta}(t)\ ,
\end{equation}
with $H$ being $d^2 \times d^2$  Hadamard matrix defined by
$$   H_{ij,kl} = \omega^{-il + jk}  \ .  $$
This definition implies that $H_{\alpha\beta}$ is a Hermitian matrix.
%%$H = F \ot F$ and $F_{kl} = \omega^{kl}$ stands for the $d \times d$  matrix of the discrete Fourier transform.
Simple algebra gives
\begin{equation}\label{Lt}
  L_t(X)=\sum_{k=1}^{d^2-1}\, \gamma_{k}(t)\, [U_{k} X U_{k}^\dag - X],
\end{equation}
where the local decoherence rates read
\begin{equation}\label{gamma}
  \gamma_\alpha(t) = \frac{1}{d^2} \sum_{\beta=0}^{d^2-1} H_{\alpha\beta}\, \mu_\beta(t)\ ,
\end{equation}
and
\begin{equation}\label{}
\mu_{\alpha}(t)=\frac{\dot{\lambda}_{\alpha}(t)}{\lambda_{\alpha}(t)}\ .
\end{equation}
Note, that the sum in (\ref{Lt}) starts from $k=1$ \cite{1} and hence there are $d^2-1$ independent decoherence rates $\gamma_k(t)$. Indeed, formula (\ref{gamma}) shows that $\gamma_0 = -  \sum_{k=1}^{d^2-1}\, \gamma_{k}$. It is therefore clear that $\Lambda_t$ defines CP-divisible dynamics iff $\gamma_k(t) \geq 0$ for all $t\geq 0$. Note, that given a map, i.e. a set of $p_\alpha(t)$, it is in general very hard to check for CP-divisibility. Conversely, given a time-local generator (\ref{Lt}) it is very hard to check whether it gives rise to a legitimate quantum evolution described by CP map $\Lambda_t$. The generator $L_t$ is legitimate iff  $p_\alpha(t) \geq 0$ for $\alpha=0,1,\ldots,d^2-1$. Using $H^{-1} = \frac{1}{d^2} H$ one easily inverts (\ref{gamma}) and finds
\begin{equation}\label{}
  p_\alpha(t) = \frac{1}{d^2} \sum_{\beta=0}^{d^2-1} H_{\alpha\beta} \, \lambda_\beta(t)\ ,
\end{equation}
where
\begin{equation}\label{}
  \lambda_\beta(t) = \exp\Big[\sum_{k=1}^{d^2-1} H_{\beta k} \Gamma_k(t)\Big] \ ,
\end{equation}
with $\Gamma_k(t) = \int_0^t \gamma_k(\tau) d\tau$. Conditions $p_\alpha(t) \geq 0$ provides highly nontrivial constraints for $\gamma_k(t)$. Note, that to have $p_\alpha(t) \geq 0$ it is sufficient $\Gamma_k(t) \geq 0$. Indeed, since $\Lambda_t = \exp[\int_0^t L_\tau d\tau]$ and
\begin{equation}\label{}
  \int_0^t L_\tau(X) d\tau = \sum_{k=1}^{d^2-1}\, \Gamma_{k}(t)\, [U_{k} X U_{k}^\dag - X],
\end{equation}
it follows that if $\Gamma_k(t) \geq 0$ then $\int_0^t L_\tau d\tau$ defines a legitimate Lindblad generator and hence $\exp[\int_0^t L_\tau d\tau]$ defines a CP-map. However, it should be stressed that  $\Gamma_k(t) \geq 0$ is not a necessary condition.

\begin{Example} For $d=2$ one has $U_k = \sigma_k$ $(k=1,2,3$), where $\sigma_k$ are Pauli matrices and hence \cite{Erika,Filip}
\begin{equation*}\label{}
  \lambda_1(t) = \exp(-2[\Gamma_2(t) + \Gamma_3(t)])\ , \ \ +\ {\rm cycl.\ perm.}
\end{equation*}
The corresponding map $\Lambda_t = \exp[\int_0^t L_\tau d\tau]$ is CP iff
\begin{equation}\label{lambda-123}
  \lambda_1(t) + \lambda_2(t) \leq 1 + \lambda_3(t)\ , \ \  +\ {\rm cycl.\ perm.}
\end{equation}
An interesting example of $\gamma_k(t)$ satisfying (\ref{lambda-123}) but violating $\Gamma_k(t) \geq 0$ was recently provided in \cite{Erika}:
\begin{equation}\label{tanh}
  \gamma_1(t) = \gamma_2(t) = \frac c2 \ , \ \ \gamma_3(t)= - \frac c2 \tanh(ct)\ ,
\end{equation}
with $c > 0$. One finds $p_3(t)=0$ and
$$ p_1(t) = p_2(t) = \frac 14 [1 - e^{-ct}]\ ,   $$
and hence the corresponding dynamical map reads
\begin{equation}\label{}
  \Lambda_t(\rho) = \frac{1+e^{-ct}}{2}\, \rho + \frac{1-e^{-ct}}{4}\, (\sigma_1 \rho \sigma_1 + \sigma_2 \rho \sigma_2) \ .
\end{equation}
Interestingly $\Lambda_t$ is a convex combination of two Markovian semigroups $\Lambda^{(1)}_t$ and $\Lambda^{(2)}_t$ generated by
\begin{equation}\label{}
  L^{(k)}_t(\rho) = \frac c2 [ \sigma_k \rho \sigma_k - \rho] \ ; \ \ k=1,2.
\end{equation}
One finds $\Lambda_t = \frac 12 ( \Lambda^{(1)}_t +\Lambda^{(2)}_t )$.

\end{Example}

\begin{Example} \label{E3} This construction may be easily generalized for $d=3$. Let us assume that
$$ \gamma_k(t)= \frac c3\ , \ \ \mathrm{for}\ k\neq 4,8\ .    $$
% = \ldots = \gamma_6(t) = 1\ . $$
Note, that $[U_4,U_8]=0$ (see Appendix for the list of $U_k$).
We look for $\gamma(t) := \gamma_4(t)=\gamma_8(t)$
such that  $p_4(t)=p_8(t)=0$.
One easily finds
\begin{equation}\label{}
  \gamma(t) = - \frac{2c}{3} \frac{e^{2ct}  - e^{-ct}}{e^{2ct}  +2 e^{-ct}}\ ,
\end{equation}
which proves that $\gamma(t) < 0$ for $t > 0$. Note that  $p_k(t)=p(t)$ ($k \neq 4,8$) with
\begin{equation}\label{}
  p(t) =\frac{1}{9}\Big(1-e^{-ct/3}\Big) \ .
\end{equation}
Similarly as for $d=2$ this evolution may be represented as a convex combination of three  Markovian semigroups $\Lambda^{(1)}_t$, $\Lambda^{(2)}_t$ $\Lambda^{(3)}_t$   generated by
\begin{eqnarray}%\label{}
  L^{(1)}_t(\rho)& = & c [ U_1 \rho U_1^\dagger +U_2 \rho U_2^\dag - 2\rho] \ , \nonumber \\
    L^{(2)}_t(\rho) &=  &c [ U_3 \rho U_3^\dagger +U_6 \rho U_6^\dag - 2\rho] \ , \\
      L^{(3)}_t(\rho)& = & c [ U_5 \rho U_5^\dagger +U_7 \rho U_7^\dag - 2\rho] \ . \nonumber
\end{eqnarray}
Note, that $[U_1,U_2]=[U_3,U_6]=[U_5,U_7]=0$.
One finds $\Lambda_t = \frac 1 3 ( \Lambda^{(1)}_t +\Lambda^{(2)}_t + \Lambda^{(3)}_t )$. Again, $\Gamma_4(t) = \Gamma_8(t)<0$ but the evolution $\Lambda_t$ is well defined. It is clear that one may generalize this example for arbitrary $d$.

\end{Example}

%\section{$k$-divisibility}

Let us observe that $L_t$ may be rewritten as follows
\begin{equation}\label{}
  L_t(X) = \Phi_t(X) + 2\gamma_0(t) X\ ,
\end{equation}
where the map $\Phi_t$ is defined via
\begin{equation}\label{}
  \Phi_t(X) = \sum_{k=1}^{d^2-1} \gamma_k(t) U_k X U_k^\dagger - \gamma_0(t)U_0 X U_0^\dagger\ ,
\end{equation}
and
$$\gamma_0(t) = -\sum_{k=1}^{d^2-1} \gamma_k(t)\ . $$
Now, the corresponding solution $V_{t,s} = \exp[ \int_s^t L_\tau d\tau]$ reads
\begin{eqnarray*}\label{}
  V_{t,s} = v(t;s) \exp\Big[\int_s^t \Phi_\tau d\tau\Big] \ ,
   % &=& v(t;s) \Big( \rho + \int_s^t \Phi_\tau d\tau \rho + \frac 12 \Big[ \int_s^t \Phi_\tau d\tau \Big]^2 \rho + \ldots \Big)
\end{eqnarray*}
where the scaling factor $v(t;s)$ is given by
\begin{equation*}\label{}
v(t;s)= \exp\Big(2   \int_s^t\gamma_0(\tau)d\tau\Big)\ .
\end{equation*}
It is therefore clear that if the map $\Phi_t$ is $k$-positive for all $t \geq 0$, then $\Lambda_t$ is $k$-divisible.

To check for $k$-divisibility we shall use the following result from \cite{CMP}: let $\Phi(X) = \sum_{\alpha=0}^{d^2-1} a_\alpha U_\alpha X U_\alpha^\dagger$ with $U_\alpha$ being Weyl unitary operators and real parameters $a_\alpha$. Clearly, if $a_\alpha \geq 0$, then $\Phi$ is CP. Suppose now that some $a_\alpha$ are negative, that is,
\begin{equation}\label{}
  \Phi(X) = \sum_{i=1}^{M} b_i U_i X U_i^\dagger   -   \sum_{j=1}^N c_j U_j X U_j^\dagger  \ ,
\end{equation}
with $M+N = d^2$ and $b_i,c_j \geq 0$ (a set $\{U_i,U_j\}$ defines a permutation of $\{U_\alpha\}$). It means that $\Phi$ is a difference of two CP maps. Let $k$ be a positive integer such that $kN < d$. One proves \cite{CMP} that if
\begin{equation}\label{bc}
  b_i \geq \frac{k}{d - kN}\, \sum_{j=1}^N c_j \ ; \ \ \ i=1,\ldots,M\ ,
\end{equation}
then $\Phi$ is $k$-positive. Moreover, if (\ref{bc}) is violated for at least one $i \in \{1,\ldots,M\}$, then $\Phi$ is not $(k+1)$-positive. Hence,  conditions (\ref{bc}) are sufficient for $k$-positivity and necessary for $(k+1)$-positivity.

Note, that if $k=1$, then $N \leq d-1$ and hence at each moment of time there are at most $d-1$ negative rates $\gamma_l(t)$. Let $N=d-1$ and suppose, that $\gamma_1(t), \ldots, \gamma_{d-1}(t) < 0$. Formula (\ref{bc}) implies
\begin{equation}\label{}
  \gamma_k(t) \geq |\gamma_1(t)| +  \ldots +  |\gamma_{d-1}(t)| \ ,
\end{equation}
or equivalently
\begin{equation}\label{}
  \gamma_k(t) + \gamma_1(t) +  \ldots +  \gamma_{d-1}(t) \geq 0 \ ,
\end{equation}
for $k=d,\ldots,d^2-1$. Replacing $\{\gamma_1(t), \ldots, \gamma_{d-1}(t)\}$ by an arbitrary set  $\{\gamma_{i_1}(t), \ldots, \gamma_{i_{d-1}}(t)\}$ one finds that if for any $d$-tuple $\{i_1,\ldots,i_{d}\} \subset \{1,2,\ldots,d^2-1\}$ the following condition is satisfied
\begin{equation}\label{P}
  \gamma_{i_1}(t) + \ldots + \gamma_{i_{d}}(t) \geq 0\ ,
\end{equation}
for all $t \geq 0$, then $\Lambda_t$ is P-divisible.

\begin{Remark} It is easy to show that random unitary evolution is P-divisible iff it satisfies the well known BLP condition \cite{BLP}:
\begin{equation}\label{blp}
  \frac{d}{dt} || \Lambda_t(\rho_1 - \rho_2)||_{\rm tr} \leq 0 \ ,
\end{equation}
for any pair of initial states $\rho_1$ and $\rho_2$. Hence,  (\ref{P}) implies (\ref{blp}).
\end{Remark}

\begin{Remark} Interestingly, if the random unitary evolution is P-divisible, then
\begin{equation}\label{S}
  \frac{d}{dt} S(\Lambda_t(\rho)) \geq 0 \ ,
\end{equation}
where $S$ denotes the von Neumann entropy. It shows that whenever the inequality (\ref{S}) is violated the evolution is essentially non-Markovian.
\end{Remark}

\begin{Remark} Authors of \cite{Pater} introduced the geometric measure of non-Markovianity {\em via}
\begin{equation}\label{}
   \mathcal{N}[\Lambda_t] = \frac{1}{V(0)} \int_{\frac{d}{dt} V(t)>0} \frac{dV(t)}{dt} dt\ ,
\end{equation}
where $V(t)$ denotes the volume of admissible states at time $t$. It is clear that for Markovian evolution one has $\frac{d}{dt} V(t)\leq 0$. Note, that
\begin{equation}\label{G}
  \sum_{k=1}^{d^2-1} \gamma_k(t) =  - \gamma_0(t) \geq 0\ ,
\end{equation}
guaranties $ \mathcal{N}[\Lambda_t] =0$. The geometric condition (\ref{G}) is much weaker than condition for P-divisibility (\ref{P}).
\end{Remark}

\begin{Example} For $d=2$ conditions (\ref{P}) give
\begin{equation*}\label{}
  \gamma_1(t) + \gamma_2(t) \geq 0 \ ,  \gamma_1(t) + \gamma_3(t) \geq 0  \ , \gamma_2(t) + \gamma_3(t) \geq 0  .
\end{equation*}
Actually, it was shown \cite{Filip} that these conditions are also necessary for P-divisibility. Note, that $\gamma_k(t)$ defined in (\ref{tanh}) satisfy these conditions and hence the corresponding dynamics is P-divisible (but not CP-divisible since $\gamma_3(t) < 0$).
\end{Example}

\begin{Example} For $d=3$ conditions (\ref{P}) give
\begin{equation}\label{Pn=3}
  \gamma_{i_1}(t) +   \gamma_{i_2}(t) +  \gamma_{i_3}(t) \geq 0\ ,
\end{equation}
for all triples $\{i_1,i_2,i_3\} \subset \{1,\ldots,8\}$. Conditions (\ref{Pn=3}) are sufficient (but not necessary) for P-divisibility. For $k=2$ one has $N \leq 1$ and hence taking $N=1$ the formula (\ref{bc}) implies: if
\begin{equation}\label{2-div}
  \gamma_{i_1}(t) +   2\gamma_{i_2}(t)  \geq 0\ ,
\end{equation}
for all pairs $\{i_1,i_2\} \subset \{1,\ldots,8\}$, then the evolution is 2-divisible. Note, that conditions (\ref{Pn=3}) are sufficient for P-divisibility and necessary for 2-divisibility whereas (\ref{2-div}) are sufficient for 2-divisibility. It is clear that (\ref{2-div}) are much stronger than (\ref{Pn=3}). Hence, if all $\gamma_k(t) \geq 0$ the evolution is Markovian and ${\rm NMD}[\Lambda_t]=0$. If $\gamma_k(t) \ngeq 0$ but condition (\ref{2-div}) is satisfied then ${\rm NMD}[\Lambda_t]=1$, that is, the evolution is non-Markovian but still 2-divisible. Finally, if (\ref{2-div}) is violated but (\ref{Pn=3}) is satisfied then ${\rm NMD}[\Lambda_t]=2$, that is, the evolution is non-Markovian but still P-divisible. However, the violation (\ref{Pn=3}) does not necessarily mean that $\Lambda_t$ is essentially non-Markovian. Actually, we conjecture that this evolution is P-divisible.  
\end{Example}

%\section{Conclusions}
To summarize: we derived a hierarchy of conditions which guarantee $k$-divisibility of the random unitary evolution of  $d$-level quantum system. It is shown how these conditions are related to well known BLP condition \cite{BLP} and the geometric condition \cite{Pater}.

\section*{Acknowledgements}

This paper was partially supported by the National Science Center project DEC-
2011/03/B/ST2/00136.

\section*{Appendix}

Weyl matrices for $d=3$: $U_0 = \mathbb{I}_3$ and
$$   U_1=\left(
\begin{array}{ccc}
 0 & 1 & 0 \\
 0 & 0 & 1 \\
 1 & 0 & 0
\end{array}
\right),\ \   U_2=\left(
\begin{array}{ccc}
 0 & 0 & 1 \\
 1 & 0 & 0 \\
 0 & 1 & 0
\end{array}
\right),$$
$$ U_3=\left(
\begin{array}{ccc}
 1 & 0 & 0 \\
 0 & \omega & 0 \\
 0 & 0 & \omega^2
\end{array}
\right),\quad U_4=\left(
\begin{array}{ccc}
 0 & 1 & 0 \\
 0 & 0 & \omega \\
 \omega^2 & 0 & 0
\end{array}
\right), $$
$$ U_5=\left(
\begin{array}{ccc}
 0 & 0 & 1 \\
 \omega & 0 & 0 \\
 0 & \omega^2 & 0
\end{array}
\right), \quad U_6=\left(
\begin{array}{ccc}
 1 & 0 & 0 \\
 0 & \omega^2 & 0 \\
 0 & 0 & \omega
\end{array}
\right),$$
$$ U_7=\left(
\begin{array}{ccc}
 0 & 1 & 0 \\
 0 & 0 & \omega^2 \\
 \omega & 0 & 0
\end{array}
\right),\quad U_8=\left(
\begin{array}{ccc}
 0 & 0 & 1 \\
 \omega^2 & 0 & 0 \\
 0 & \omega & 0
\end{array}
\right) \ ,$$
with $\omega = e^{2\pi i/3}$ and $\omega^2 = \omega^* =  e^{-2\pi i/3}$.

%===========================      BIBLIOGRAPHY    ====================================


\begin{thebibliography}{1}
\bibitem{k-pos} A map $\Phi : \mathcal{B}(\mathcal{H}) \rightarrow   \mathcal{B}(\mathcal{H})$ is $k$-positive iff $\oper_k \ot \Phi$ is positive, where $\oper_k$ denotes identity map acting in the space of $k \times k$ complex matrices.
\end{thebibliography}
\end{document}